# Demonstration of 50 km Fiber-optic two-way quantum time transfer at femtosecond-scale precision

Huibo Hong[1,2], Runai Quan[1,2,*], Xiao Xiang[1,2], Wenxiang Xue[1,2], Honglei Quan[1,2], Wenyu Zhao[1,2], Yuting Liu[1,2], Mingtao Cao[1,2], Tao Liu[1,2], Shougang Zhang[1,2], Ruifang Dong[1,2,*]

[1] Key Laboratory of Time and Frequency Primary Standards, National Time Service Center, Chinese Academy of Sciences, Xi'an, 710600, China

[2] School of Astronomy and Space Science, University of Chinese Academy of Sciences, Beijing, 100049, China

[*]E-mail: quanrunai@ntsc.ac.cn

[*]E-mail: dongruifang@ntsc.ac.cn

**Abstract:** The two-way quantum time transfer method has been proposed and experimentally demonstrated for its potential enhancements in precision and better guarantee of security. To further testify its advantage in practical applications, the applicable direct transmission distance as well as the achievable synchronization precision between independent time scales is of great interest. In this paper, an experiment on two-way quantum time transfer has been carried out over a 50 km long fiber link. With the common clock reference, a short-term stability of 2.6 ps at an averaging time of 7 s and a long-term stability of 54.6 fs at 57300 s were obtained. With independent clock references, assisted by microwave frequency transfer technology, the achieved synchronization showed almost equal performance and reached a stability of 89.5 fs at 57300 s. Furthermore, the spectral consistency of the utilized entangled photon pair sources has been studied concerning its effect on the transfer accuracy and long-term stability. The results obtained have promised a bright future of the two-way quantum time transfer for realizing high-precision time synchronization on metropolitan area fiber links.

## Introduction:

High-precision time synchronization is of great importance in wide variety of fields such as [1], geodesy [2, 3], astronomy [4], deep space exploration [5], communication [6, 7], and scientific research and measurements [8], etc. Among different methods, the two-way time transfer is considered an effective way to synchronize two remote time scales with high precision and accuracy independently on the variations of the interconnecting channel [9]. In practical applications, traditional transfer schemes based on satellites are no longer sufficient to meet the requirements of users in terms of accuracy, stability, and reliability [10, 11]. Fiber-optic time transfer techniques have become widespread both in scientific and commercial applications based on its ease to use and inherent symmetry of the transmission medium. Over fiber lengths of hundreds of kilometers [12-14], precisions of tens of picoseconds have been reported, marking a significant improvement over satellite-based two-way time transfer techniques [15, 16]. Further cooperating with quantum technology, the quantum time transfer method has shown great potential in enhancing the precision [17-19] and providing better guarantee of security [20]. As a case in point, we recently reported a two-way quantum time transfer experiment over a 20-km optical fiber link [21]. The experimental results showed a minimum time stability of 45 fs and an accuracy of 2.46 ps in terms of the fiber length, which demonstrates an appreciable improvement over the classical counterparts. To showcase its apparent advantage in practical applications, the applicable direct transmission distance as well as the achievable synchronization precision between independent time scales is of valuable interest.

In this paper, we further implement the two-way quantum time transfer experiment with the fiber distance extensively extended to 50 km. By referenced to the same clock, the time transfer stability has been achieved to be 2.6 ps at an averaging time of 7 s and 54.6 fs at 57300 s, which are equivalent with the results given in [21]. Based on this quantum time transfer system, the time synchronization between two independent clock references has also been investigated. Assisted with the microwave frequency transfer technology to build up a common frequency reference, the time synchronization has achieved an almost equally good performance as the case of common clock reference and reached a stability of 89.5 fs at 57300 s. The spectral consistency of the utilized entangled photon pair sources has been subsequently studied concerning its effect on the transfer accuracy. The agreement between theory and experimental results verifies that, the accuracy is linearly proportional to the fiber length due to the center-wavelength difference between the forward and backward transmitted photons enlarged by the dispersion, which is analogous to its classical counterpart [22]. It has also been verified that, the long-term stability can be improved by using entangled photon pair sources with better spectral consistency to reduce the wavelength-dependent ambient condition (mainly temperature) induced variations [23]. The achieved results in the experiment have promised a bright future of the two-way quantum time transfer for realizing high-precision time synchronization on metropolitan area fiber links.

**Experimental setup：**

The schematic diagram of the fiber-optic two-way quantum time transfer (F-TWQTT) experimental setup is shown in Fig. 1. The two time scales to be synchronized, Clock A and Clock B, are supposed to be located at site A and site B, which is linked by a 50 km fiber link. Both sites are equipped with a telecommunication-wavelength frequency entangled photon-pair source, two single-photon detectors (SPDs), and an Event timer (ET) which is referenced to its local time scale. Via the spontaneous parametric down conversion process (SPDC), each photon-pair source, denoted as Source A and Source B, is generated from a 10mm-long, type-II quasi-phase-matched periodically poled lithium niobate (PPLN) waveguide pumped by a 780-nm quasi-monochromatic laser [24]. After filtering out the residual pump, the frequency anti-correlated entangled photon pairs with orthogonal polarizations (denoted as signal and idler photons) are coupled into a fiber polarization beam splitter (FPBS) for spatial separation and subsequent distribution. The SPDs are superconductive nanowire single photon detectors (SNSPDs, Photec Ltd.), with a FWHM timing jitter about 68 ps and a detection efficiency of 65% [25, 26]. The idler photons ($i_1$) from Source A are locally detected by the SNSPD D1 at site A, while the signal photons ($s_1$) are sent forward through the 50 km fiber link to site B and detected by the SNSPD D4. Likewise, the idler photons ($i_2$) from Source B are locally detected by the SNSPD D2 at site B, while the signal photons ($s_2$) are sent backward to site A and detected by the SNSPD D3. Via the optical circulators (OC1&OC2), the forward and backward transmitted photons share the same fiber link. According to nonlocal dispersion cancellation effect [20], fiber Bragg gratings (FBG1&FBG2) with a dispersion of 825 ps/nm (DCMCB-SN-050P1FA, Proximion Inc) have been inserted into the signal arms to compensate for the dispersion experienced by idler photons in the 50 km fiber link. Commercial event timers (A033ET/USB, Eventech Ltd) ET A and ET B are used to record the arrival times of the detected photons. ET A at site A records the arrival times at D1 and D3 as $\{t_1^{(j)}\}$ and $\{t_3^{(j)}\}$ respectively, where the subscript $j = 1 \dots n$ denotes the $j$-th time tag of the recorded time sequences. Accordingly, ET B at site B records the arrival times at D2 and D4 as $\{t_2^{(j)}\}$ and $\{t_4^{(j)}\}$. By applying the nonlocal coincidence identification algorithm onto the time sequences [27], the time differences $t_4 - t_1$ and $t_3 - t_2$ can be extracted. Under the assumption that the two-way transfer setup is symmetric in both directions and the FBGs introduce the same delay, the time offset between the two clocks is then given by $t_0 = ((t_4 - t_1) - (t_3 - t_2))/2$.

Utilizing a fiber-based microwave frequency transfer device, the 10 MHz frequency signal of the time scale at site A is also transferred through another 50 km fiber to site B to provide an auxiliary 10 MHz frequency reference for ET B.

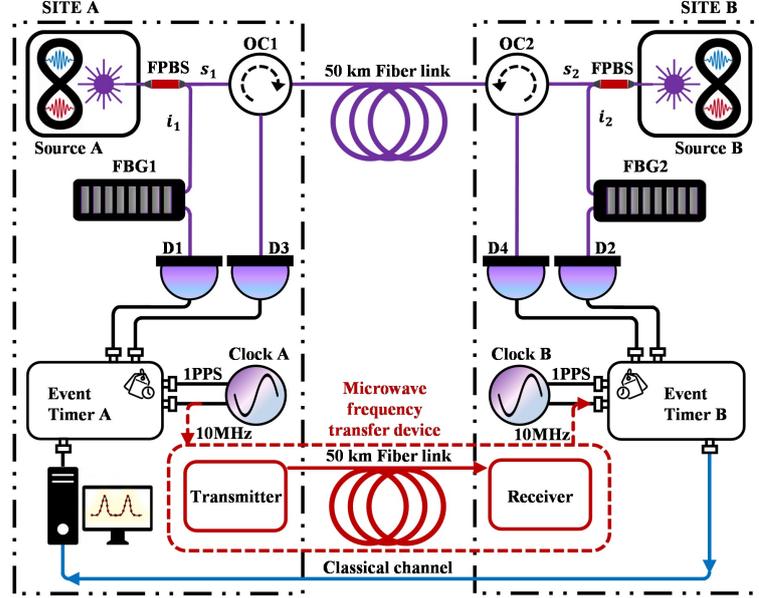

Fig. 1. Schematic diagram of the F-TWQTT experimental setup. At site A (B), the idler photons from Source A (B) are kept locally and detected by the SNSPD D1 (D2) while the signal photons are sent through the 50 km fiber link to site B (A) and detected by the SNSPD D4 (D3).The optical circulators (OC1&OC2) are used to forward and backward transmit photons in the same fiber. The fiber Bragg grating modules (FBG1&FBG2) are for compensating the dispersion in the 50 km fiber link nonlocally. By referenced to the local time scales, ET A and ET B record the arrival times of the detected photons. In addition, the microwave frequency transfer of the 10MHz from site A to B via another independent 50 km fiber link is sketched in red dashed frame.

**Results and Analysis：**

With the lab-own H-maser as the common reference of the two sites, the time transfer performance over the 50 km fiber link was firstly investigated. Due to the sampling rate limitation of the event timers, the photon count rate of each SNSPD was set around 20 kHz. By recording the arrival times in 2.5 s, each registered time sequence contained about 50000 time stamps. As the ETs need additional time for data saving and operation recovery, it took around 7 seconds for each measurement run. From these recorded time sequences, the temporal coincidence distribution histograms for the differences of $t_4 - t_1$ and $t_3 - t_2$ were built through the nonlocal coincidence identification algorithm, the similar coincidence widths of 120 ps in FWHM showed that the dispersions experienced by the signal photons in the 50 km fiber link have been effectively cancelled by the FBGs inserted the idler arms [Li et al., PRA 2019]. From the Gaussian-fitted peaks of $t_4 - t_1$ and $t_3 - t_2$, the result of the time offset $t_0$ was given. The corresponding time transfer stability, in terms of time deviation (TDEV), versus the averaging time is shown in Fig. 2 by red circles, which achieves 2.6 ps at an averaging time of 7 s and reaches a minimum of 54.6 fs at 57300 s.

Based on this quantum time transfer setup, the time synchronization experiment between two independent time scales were implemented by using a commercial Rb clock (PRS10, SRS. Inc) as the reference of ET A. The achieved time stability in terms of TDEV is shown in Fig. 2 by dark cyan squares, and a value of 9.9 ps at 6 s averaging time is achieved, which has a corresponding result of 2.9E-12 in Allan deviation (ADEV). The agreement of this ADEV result with the intrinsic ADEV of the Rb clock (< 3E-12) indicates that the time synchronization stability is ultimately determined by the performance of the less-stable reference. As the averaging time increases, the long-term stabilities for the case of independent reference clocks are much worse than that with common reference clock. Therefore, establishing a common frequency reference is a promising method for significantly improving the quantum clock synchronization performance, as has been stated before [28].

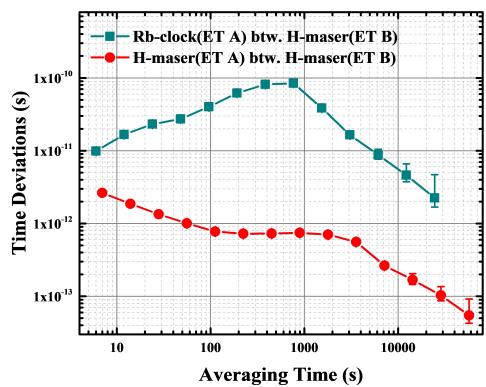

Figure 2. The measured TDEV results of time offset $t_0$ between the two sites over the 50 km fiber link after implementing the F-TWQTT protocol for the case (a) with both ET A and B being referenced to the H-maser (in red circles) and (b) with ET A being referenced to the Rb clock while ET B to the H-maser (in dark cyan squares).

Subsequently, the improved quantum time synchronization experiment was implemented between two independent time scales by applying the microwave frequency transfer technology [29]. With a stability of 1.15E-14/s and 3.2E-16/40000s in ADEV (as shown in Fig. 3 by blue triangles), the 10 MHz frequency signal of the Rb clock was transferred through another parallel 50 km fiber link to site B, which was used as the 10 MHz frequency for ET B. Based on this configuration, the TDEV was achieved as 2.7 ps at an averaging time of 7 s and reached 89.5 fs at 57300 s. The results are given in Fig. 3 by black squares. For comparison, the measured TDEVs for the case with both ET A and B being referenced to the H-maser is also shown in Fig. 3 by red circles. As can be seen that, the synchronization stability improved by the microwave frequency transfer technology can achieve almost equally good performance as the case with common time scale reference both in short-term and long-term stability. Nonetheless, within the averaging time range between 100 s and 2000 s, there exists an apparent bump. By looking at the transferred stability of the 10MHz reference frequency (shown by blue triangles), the similar bump was observed, which has been discussed in the previous literature and attributed to the periodical temperature variation in the laboratory [29]. Therefore, the bump in the TDEV curve reflects the influence of the frequency transfer performance on the achieved time synchronization stability.

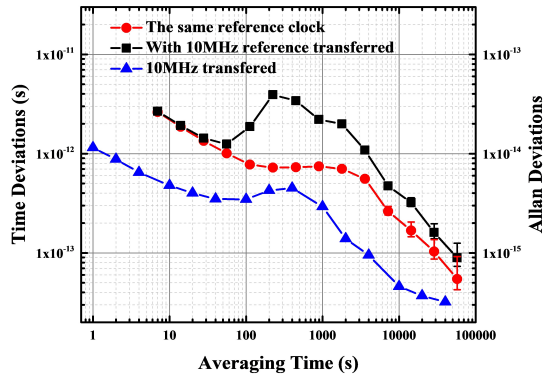

Figure 3. The measured TDEVs of time offset $t_0$ over the 50 km fiber link for the case (a) with the H-maser as the common time scale reference (in red circles) and (b) with independent time scale references but sharing the same 10 MHz frequency reference from the Rb clock via the fiber-optic microwave frequency transfer technology (in black squares). The frequency transfer stability in terms of ADEV of the 10 MHz frequency signal is also given by blue triangles.

The time transfer accuracy of the system with respect to systematic parameters, such as the spectral consistency of the utilized entangled photon pair sources and fiber distance, was also investigated. By adjusting the two photon pair sources in the setup to satisfy two cases of spectral consistency as shown in Table 1, which were prepared via fine adjustment of the pump wavelengths and the temperature of the PPLN waveguides, the absolute time offset results ($\bar{t}_0$) with and without the 50 km fiber in the setup were measured and compared. For the case of low spectral consistency, the signal photon of Source A was measured having a center wavelength of 1561.24 nm and a bandwidth of 3.55 nm, while that of Source B was measured having a center wavelength of 1560.54 nm and a bandwidth of 3.5 nm. The wavelength difference between the two signal photon beams was thus 0.7 nm. According to Ref. [22], it will introduce an extra time offset ($\tau'$) that cannot be cancelled by the two-way transfer setup:

$$\tau' = LD(\bar{\lambda}_{s,A} - \bar{\lambda}_{s,B}) \qquad (1)$$

where $L$ is the fiber length, $D$ is the dispersion of the fiber which is about 17 ps/nm/km for the single mode fiber [30], and $\bar{\lambda}_{s,A(B)}$ is the center wavelength of the signal photons of Source A (B). By substituting the above parameters of $\bar{\lambda}_{s,A}$=1561.24 nm, $\bar{\lambda}_{s,B}$=1560.54 nm into Eqn. (1), an extra time bias of 595 ps was expected. The corresponding experimental test was implemented by comparing the measured absolute time offset values between the conditions with and without the 50km fiber in the setup, each of which was obtained by averaging 36 sets of time offset measurements. As shown in Fig. 4, a bias of 524.7±25.3 ps was measured, which agrees well with the theoretical expectation. It can be seen the variation of the measured time offset values with the 50km fiber in the setup is one magnitude lower than that without the 50km fiber in. It is due to the dispersion cancellation effect introduced by the embedded FBGs in the idler arms. For the case with high spectral consistency between Source A and Source B, $\bar{\lambda}_{s,A}$=1560.26 nm and $\bar{\lambda}_{s,B}$=1560.24 nm was achieved and an extra time bias of 17 ps was accordingly estimated. Based on the experimental test, the extra time bias was measured to be 4.3±35.5 ps, which was also shown in Fig. 4. The good agreements between the experimental results and theoretical simulations for both cases showed that, analogous to its classical counterpart, the spectral consistency of the utilized frequency entangled photon pair sources makes a nontrivial effect on the achievable synchronization accuracy of the two-way quantum time transfer system.

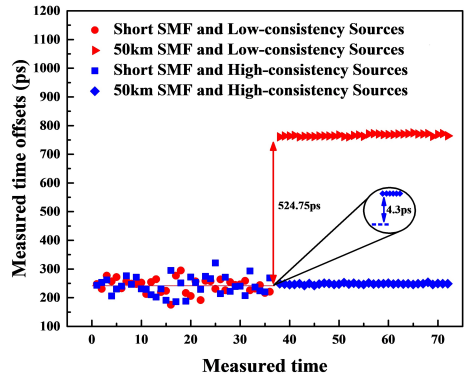

Figure 4. Comparison of the measured absolute time offset values between the conditions with and without the 50km fiber in the setup. Two cases, with respect to low-consistency sources (red) and high-consistency sources (blue), are considered.

**Table 1.** Tabulation of the experimentally measured and theoretically simulated time offset bias over 50 km fiber links in terms of the low-consistency sources and high-consistency sources.

|  | $\bar{\lambda}_{s,A}$ | $\bar{\lambda}_{s,B}$ | $\Delta\bar{\lambda}_S$ | $\tau'_{theo}$ | $\tau'_{exp}$ |
|---|---|---|---|---|---|
| Low consistency | 1561.24 nm $\pm$0.038 nm | 1560.54 nm $\pm$0.103 nm | 0.70 nm $\pm$0.11 nm | 595ps $\pm$93.3ps | 524.7 ps $\pm$25.3 ps |
| High consistency | 1560.26 nm $\pm$0.037 nm | 1560.24 nm $\pm$0.044 nm | 0.02 nm $\pm$0.06 nm | 17ps $\pm$48.9ps | 4.3 ps $\pm$35.5 ps |

The enhancement impact of the spectral consistency between the two entangled sources on the synchronization stability was also tested. With a common clock reference, Fig. 5 depicts the two TDEV curves corresponding to the high-consistency (in red circles) and low-consistency (in royal squares) cases respectively. It can be clearly seen that, the TDEV curve of the high-consistency case shows an apparent lower level than that of the low-consistency case over the averaging time range of 2000-50000 s. Therefore, by using entangled photon pair sources with better spectral consistency, the stability can be improved, especially in the long term. This result verifies the reduction of the wavelength-dependent ambient condition (mainly temperature) induced variations [23] is also necessary to improve the performance of two-way quantum time transfer system.

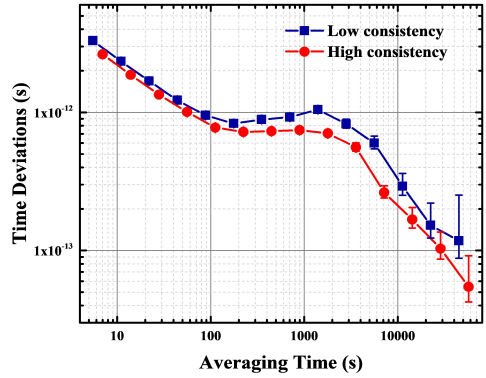

Figure 5. The TDEV over 50 km fiber links with high-consistency sources (in red circles) and low-consistency sources (in royal squares)

## Conclusion

In this paper, the two-way quantum time synchronization experiment with a femtosecond-scale precision was demonstrated over a 50 km fiber-optic link, which is close to the distance between metropolitan optical fiber network nodes. With the common clock reference, the time transfer stability in TDEV was achieved as 2.6 ps at an averaging time of 7 s, with the long-term stability reaching 54.6 fs at 57300 s. With independent clock references, incorporating with the microwave frequency transfer technology, the time stability showed almost equally good performance as the case for common clock reference, and a long-term stability of 89.5 fs at 57300 s was achieved. The results could be further improved by using new ETs with higher acquisition rates, applying new SNSPDs with lower timing jitters and optimizing the nonlocal dispersion effect. Investigations were also made concerning the effect of the spectral consistency of the utilized entangled photon pair sources on the system. It has been shown that, analogous to the classical two-way system, the optimization of the spectral consistency of the entangled light sources at the two sites is necessary to improve both the time synchronization accuracy and the long-term stability. The excellent synchronization performance demonstrated here shows the bright prospect of the two-way quantum time transfer method in the real-field metropolitan-area time transfer systems for much higher stability.

**Acknowledgments**

This work was supported by the National Natural Science Foundation of China (Grant Nos. 12033007, 61801458, 61875205, 91836301), the Frontier Science Key Research Project of Chinese Academy of Sciences (Grant No. QYZDB-SW-SLH007), the Strategic Priority Research Program of CAS (Grant No. XDC07020200), the Youth Innovation Promotion Association, CAS (Grant No. 2021408), the Western Young Scholar Project of CAS (Grant Nos. XAB2019B17 and XAB2019B15), the Chinese Academy of Sciences Key Project (Grant No. ZDRW-KT-2019-1-0103).

**Data Availability Statement**

The data that support the fundings of this study are available from the corresponding authors upon reasonable request.